%% file: main.tex
\documentclass[conference]{IEEEtran}
\IEEEoverridecommandlockouts
\usepackage{cite}
\usepackage{amsmath,amssymb,amsfonts}
\usepackage{algorithm}
\usepackage[noend]{algpseudocode}
\usepackage{graphicx}
\usepackage{textcomp}
\usepackage{xcolor}
\usepackage{booktabs,array}
\usepackage{placeins}
\definecolor{algcomment}{RGB}{0,115,64}
\algrenewcommand{\algorithmicrequire}{\textbf{Input:}}
\algrenewcommand{\algorithmicensure}{\textbf{Output:}}
\algrenewcommand{\algorithmicindent}{1em}
\algrenewcommand{\algorithmiccomment}[1]{\hfill{\color{algcomment}\# #1}}
\newcommand{\AlgNote}[1]{\State {\color{algcomment}\# #1}}
\newcommand{\AlgEvent}[1]{\State $\triangleright$ \textbf{Event: #1}}
\usepackage[hidelinks]{hyperref}
\newif\ifdraftnotes
\draftnotestrue

\newcommand{\IACCsectionend}{}
\def\BibTeX{{\rm B\kern-.05em{\sc i\kern-.025em b}\kern-.08em
    T\kern-.1667em\lower.7ex\hbox{E}\kern-.125emX}}

\begin{document}
\title{Inference-Aware Congestion Control for Distributed MoE Inference}
\author{
\IEEEauthorblockN{Weiye Chen\IEEEauthorrefmark{1}\IEEEauthorrefmark{2},
Qi Xin\IEEEauthorrefmark{1},
Ahmed M. Abdelmoniem\IEEEauthorrefmark{2}}
\IEEEauthorblockA{\IEEEauthorrefmark{1}Beijing University of Posts and Telecommunications, Beijing, China}
\IEEEauthorblockA{\IEEEauthorrefmark{2}Queen Mary University of London, London, United Kingdom}
}
\maketitle
\input{sections/00_abstract}
\input{sections/01_introduction}
\input{sections/02_background}
\input{sections/03_design}
\input{sections/05_evaluation}
\input{sections/06_related_work}
\input{sections/07_conclusion}
\bibliographystyle{IEEEtran}
\bibliography{references}
\end{document}

%% file: sections/00_abstract.tex
\begin{abstract}
Congestion control is central to low-latency datacenter communication, yet link utilization and queue state do not reveal whether delivered bytes advance an application. This distinction matters in distributed mixture-of-experts (MoE) inference: collective transfers jointly satisfy GPU dependencies, while computation may overlap with communication. The transfer limiting a consumer can therefore change over time, and extra bandwidth may have little effect while that consumer is still computing. We study this interaction and present IACC, which combines ACK/INT feedback with collective and consumer progress when allocating QP rates. A fixed-snapshot analysis characterizes budget conservation and weighted max-min allocation. Across five trace-guided workloads at baseline arrival density, IACC lowers mean TTFT and TPOT relative to HPCC by 6.32\%--19.41\% and 7.02\%--27.33\%, respectively. Across tested workload--density settings through $1.25\times$, mean and p95 TTFT are also lower.
\end{abstract}

%% file: sections/01_introduction.tex
\section{Introduction}
\label{sec:introduction}

End-to-end request latency is a central quality-of-service (QoS) concern for distributed inference. In mixture-of-experts (MoE) serving, requests repeatedly exchange data among GPUs to dispatch tokens and combine expert outputs. These transfers carry inputs for subsequent computation, so communication delays can extend request completion time. Existing work improves MoE communication, expert execution, and serving organization~\cite{deepspeed_moe,fastermoe,lina,megascale}. As concurrent requests share network paths, bandwidth allocation becomes an important part of improving inference latency.

\input{figures/concepts_20261124/motivation}

Congestion control regulates how requests share the network. Data-center transports use ECN, delay, or in-network telemetry to respond to congestion~\cite{dctcp,dcqcn,timely,swift,hpcc}. Translating network observations into better request performance also requires knowing which computation depends on each transfer. Two allocations can keep the same bottleneck busy yet produce different GPU waiting and request completion times. Which unfinished communication should receive service now to advance inference execution?

Two properties determine the answer. First, consumer progress depends on related transfers. A collective can have several completed members while a straggling participant still withholds a required input. Accelerating a QP advances execution when it brings that dependency closer to satisfaction. Coflow scheduling captures shared completion objectives among related flows~\cite{coflow,varys,aalo}; GPU dependencies further identify which unfinished transfers currently gate computation.

Second, communication urgency changes with consumer progress. A GPU that has completed its preceding work can start when its last required input arrives. A competing consumer may still have computation that covers additional communication delay. Shifting part of its bandwidth to the blocking transfer can advance the first consumer while retaining the second consumer's start time (Fig.~\ref{fig:motivation}). Collective dependencies identify the work to advance; consumer readiness reveals its urgency and the slack available for reallocation.

IACC combines these signals in the ACK/INT feedback path. ACK/INT observations determine effective service budgets on shared directed links, while runtime computation-progress and dependency hints guide budget sharing across collectives and QP weights within each collective. Groups with computation slack yield service to groups with blocked consumers, and the feedback path directly applies the resulting absolute QP rates.

Within each connected component, IACC selects a consistent priority group so competing links advance the same dependency. Constrained progressive filling converts group budgets into rates that respect path and endpoint limits. ACK/INT observations refresh shared constraints; cached rates serve traffic between refreshes.

This paper makes four contributions:
\begin{itemize}
\item We explain why flow-level service can diverge from request-level progress: collective dependencies identify which transfers gate execution, while consumer readiness determines their urgency.
\item We design IACC, which combines network telemetry and inference state in the ACK/INT feedback path and directly applies the resulting QP rates.
\item We analyze budget conservation and weighted max-min QP allocation for a fixed snapshot, including algorithm termination and complexity.
\item We evaluate five trace-guided workloads across baseline and increased arrival densities, measuring TTFT, TPOT, RCT, GPU waiting, and utilization. Consumer-state ablation and controlled rank skew further examine how execution progress affects the results. At baseline, IACC reduces mean TTFT, TPOT, and RCT by 6.32\%--19.41\%, 7.02\%--27.33\%, and 6.45\%--21.47\%, respectively, relative to HPCC; mean and p95 TTFT are lower in all 30 tested settings through $1.25\times$.
\end{itemize}

%% file: figures/concepts_20261124/motivation.tex
\begin{figure}[!t]
\centering
\includegraphics[width=\columnwidth]{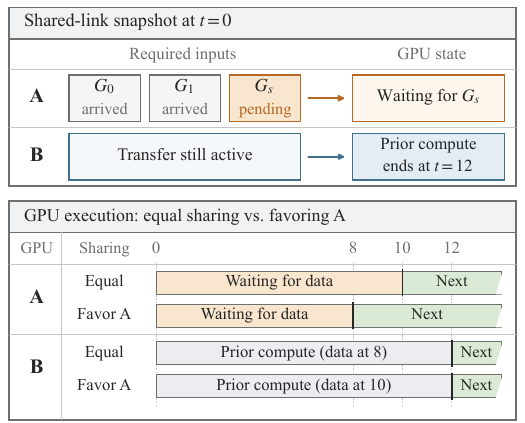}
\caption{Communication service and GPU progress. A and B share a unit-capacity link with 6 and 4 units left. Favor A uses $3{:}1$ shares while both transfers are active. A starts its next computation at $t=8$ instead of 10; B's input arrives later but its next computation still starts at $t=12$.}
\label{fig:motivation}
\end{figure}

%% file: sections/02_background.tex
\section{Background and Motivation}
\label{sec:background}

This section describes communication dependencies in MoE inference and their interaction with GPU execution. A shared-link example then motivates bandwidth allocation that considers both straggling transfers and consumer progress.

\input{figures/concepts_20261124/overview}

\subsection{Communication and Computation in MoE Inference}

Mixture-of-experts (MoE) models select a subset of experts for each token. Expert parallelism (EP) distributes experts across GPUs. An MoE layer dispatches tokens to their assigned experts, executes expert computation, and returns and combines the results. Communication thus connects successive computation stages and affects when subsequent computation can begin~\cite{lina,megascale}.

Each EP collective involves multiple ranks and transfers over multiple queue pairs (QPs). Input-dependent routing creates differences in communication volume and computational load, while concurrent collectives may compete for shared links~\cite{lina}. Their transfers can differ in path and completion progress. Meanwhile, each GPU's execution queue and dependencies determine when it can consume the results.

\subsection{From QP Progress to Collective Progress}

RDMA congestion control adjusts QP sending rates using network feedback. For example, HPCC uses in-network telemetry (INT) carried by ACKs to track link load and queue state~\cite{hpcc}. These signals describe network conditions; collective progress also depends on which required transfers remain incomplete.

A consumer requiring several inputs cannot proceed until all arrive. A straggling transfer can therefore determine when its communication dependency is satisfied, even if other transfers finish earlier. Coflow research likewise identifies coupled completion objectives among flows serving one application stage~\cite{varys}. Collective membership and remaining work reveal which transfers limit progress. Their path bottlenecks also matter: additional service on one path cannot eliminate a communication tail on another.

\subsection{From Collective Progress to GPU Computation}

Consider a computation on GPU $G_i$ that consumes communication results. Under a given execution order, let $R_i$ be the time when all prerequisites other than the target communication are satisfied, and $F_i$ the time when its required transfers finish. The earliest start time and communication-induced waiting time are

\begin{equation}
S_i = \max(R_i,F_i),
\end{equation}

\begin{equation}
W_i = \max(0,F_i-R_i).
\end{equation}

When $F_i>R_i$, unfinished communication directly delays computation, so earlier completion reduces GPU waiting. When $F_i\le R_i$, communication is covered by preceding work; completing it earlier does not advance the consumer. This preceding computation need not start at the same time as the target communication.

Remaining communication work determines how much network service is needed. Consumer readiness determines its urgency: a GPU waiting for data needs timely service, whereas ongoing preceding computation may leave slack for delaying communication.

\subsection{Bandwidth Allocation with Dependencies and Progress}

Figure~\ref{fig:motivation} illustrates concurrent collectives $A$ and $B$ sharing a directed bottleneck link $e$. Computation $C_A$ on GPU $G_c$ requires inputs from $G_0$, $G_1$, and $G_s$. The first two inputs have arrived, but the transfer from $G_s$ remains incomplete. Since $G_c$ has finished its preceding computation, it is waiting for this final input. A required transfer of $B$ competes with $G_s\to G_c$ on $e$. However, $B$'s consumer still has preceding computation in progress, and its remaining communication can finish before that computation ends under the current allocation.

The completed inputs from $G_0$ and $G_1$ cannot release $C_A$: the input from $G_s$ remains the blocking dependency. Allocating additional bandwidth to $G_s\to G_c$ can therefore shorten GPU waiting. Reassigning part of $B$'s bandwidth share to this transfer advances $C_A$ while allowing $B$ to continue transmitting. Although $B$ finishes communicating later, its consumer starts at the same time if the transfer still completes before the preceding computation ends. The feasible reallocation depends on $B$'s remaining data, computation slack, and path capacities.

Both allocations can fully utilize the link, yet produce different GPU waiting times. Collective dependencies identify the transfer that limits progress; consumer state reveals the urgency and slack that guide bandwidth allocation.

IACC combines these signals within the ACK/INT feedback path. The runtime supplies computation-progress and dependency hints. Together with network state and remaining collective work, these hints guide per-group bandwidth budgets on shared directed links and their allocation to QPs, whose sending rates are updated directly. The next section presents the design.\IACCsectionend

%% file: figures/concepts_20261124/overview.tex
\begin{figure*}[!t]
\centering
\includegraphics[width=\textwidth]{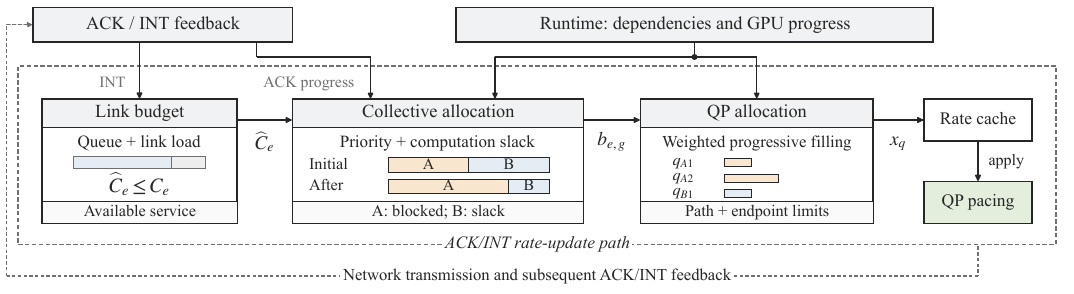}
\caption{IACC overview. ACK/INT observations determine link-service budgets. Collective dependencies and GPU progress guide collective budgets and QP weights; weighted progressive filling enforces the allocation constraints. Rates are cached and applied directly within the ACK/INT rate-update path.}
\label{fig:iaccoverview}
\end{figure*}

%% file: sections/03_design.tex
\section{IACC Design}
\label{sec:design}

IACC turns network feedback and inference progress into one QP rate decision (Fig.~\ref{fig:iaccoverview}). The ACK/INT path first determines the service available on each directed link, then distributes it using collective dependencies and consumer readiness, and finally applies the resulting rates. The runtime supplies progress and dependency records. We present the objective, network budgets, collective allocation, and rate execution in that order.

\subsection{Model and Allocation Objective}
\label{sec:overview}

At a refresh time $t$, let $\mathcal Q$ contain admitted QPs with unsent data and fresh delivered records. QP $q$ belongs to collective $g(q)$, traverses directed path $\mathcal P_q$, and has endpoint cap $P_q$. Collectives are identified by request and communication-operation IDs; a flow without collective metadata forms its own group. Define $\mathcal Q_{e,g}=\{q\in\mathcal Q:g(q)=g,e\in\mathcal P_q\}$ and $\mathcal G_e=\{g:\mathcal Q_{e,g}\ne\varnothing\}$. Given link capacity $C_e$, feedback-derived budget $\widehat C_e(t)$, and group budgets $\mathbf b$, feasible QP targets satisfy
\begin{align}
0\le x_q&\le P_q,\qquad q\in\mathcal Q,\label{eq:endpoint}\\
\sum_{q\in\mathcal Q_{e,g}}x_q&\le b_{e,g},\qquad
\sum_{g\in\mathcal G_e}b_{e,g}\le\widehat C_e(t)\le C_e.
\label{eq:linkconstraints}
\end{align}
Denote this feasible set by $\mathcal X(\mathbf b)$.

Let $F_{g,i}$ be the completion time of communication required by consumer $G_i$, and $R_{g,i}$ its readiness excluding that communication. As in Section~\ref{sec:background}, communication-induced waiting is $W_{g,i}=[F_{g,i}-R_{g,i}]^+$. Holding other dependencies fixed, advancing communication by $\epsilon\ge0$ reduces waiting by
\begin{equation}
W_{g,i}(F_{g,i})-W_{g,i}(F_{g,i}-\epsilon)
=\min\{\epsilon,[F_{g,i}-R_{g,i}]^+\}.
\label{eq:waitinggain}
\end{equation}
Here $[z]^+=\max(0,z)$. Service therefore has immediate value when it advances a blocking dependency; communication covered by computation has slack. IACC uses this distinction to set collective budgets, then computes QP rates within them.

\begin{table}[t]
\centering
\caption{State and decision variables used by IACC.}
\label{tab:notation}
\begin{tabular}{@{}p{0.23\columnwidth}p{0.70\columnwidth}@{}}
\toprule
Symbol & Meaning \\
\midrule
$u_q,o_q$ & Unsent and unacknowledged bytes of QP $q$ \\
$\mathcal{B}_g$ & Consumers currently blocked on collective $g$ \\
$R_{g,i},R_g$ & Consumer and earliest group readiness times \\
$L_{e,g},D_g$ & Remaining wire bits and estimated drain time \\
$\widetilde b_{e,g},b_{e,g}$ & Budgets before and after slack reassignment \\
$x_q,\bar x_q$ & Continuous target and quantized rate grant \\
$Q_e,\nu_e,H_e$ & Queue bytes, observed rate, and drain horizon \\
$C_e^{\mathrm{eff}},\widehat C_e$ & Effective capacity and allocatable link budget \\
\bottomrule
\end{tabular}
\end{table}

\subsection{Network-Conditioned Link Budgets}
\label{sec:networkbudget}

ACK-carried INT supplies queue bytes $Q_e$, transmitted-byte increments $\Delta B_e$, and sampling intervals $\Delta t_e$. Per-QP histories feed a shared directed-edge record; the first sample establishes a baseline. A sample requires a unique edge match on the QP path, a positive interval, and a preceding ACK within horizon $H_e=\max\{H_0,4\operatorname{RTT}_q^{\mathrm{base}}\}$. With time in seconds and rates in bits/s, define
\begin{equation}
\nu_e=\frac{8\Delta B_e}{\Delta t_e},\qquad
u_e=\max\{1,\nu_e/C_e\},\qquad
z_e=\frac{8Q_e}{C_eH_e}.
\label{eq:networkpressure}
\end{equation}
The load term $u_e$ and normalized queued work $z_e$ determine effective capacity and allocatable budget:
\begin{equation}
C_e^{\mathrm{eff}}=\max\left\{r_{\min},
\left\lfloor\frac{C_e}{u_e+z_e}\right\rfloor\right\},\qquad
\widehat C_e=\eta C_e^{\mathrm{eff}},
\label{eq:effectivecapacity}
\end{equation}
where $r_{\min}=1$ bit/s and $0<\eta\le1$ is the utilization fraction. Edge records expire after $T_{\mathrm{obs}}$; missing or stale observations yield $C_e^{\mathrm{eff}}=C_e$. The observation lifetime and drain horizon serve different purposes: one limits sample age, while the other normalizes queued work.

For $C_e\ge r_{\min}$, the denominator ensures $C_e^{\mathrm{eff}}\le C_e$. Increasing queue or load terms cannot increase the budget at a fixed horizon. This budget enters the drain estimate and all subsequent allocation constraints, coupling congestion response to the service available for inference.

\subsection{Collective Coordination and Computation Slack}
\label{sec:groupallocation}
\label{sec:tracking}

\begin{algorithm}[t]
\caption{Compute collective budgets}
\label{alg:budgets}
\small
\begin{algorithmic}[1]
\Require $\{\widehat C_e,L_{e,g},V_g,R_g,\mathcal{B}_g\}$, $t$; histories $s,c$
\State Compute $D_g$ and $\sigma_g$ using (\ref{eq:drain})--(\ref{eq:score})
\AlgNote{Select a consistent priority group}
\ForAll{contention components $K$}
  \State $U_K\gets\{g\in K: \mathcal{B}_g\ne\varnothing\}$
  \If{$c_K\notin K$} \State $c_K\gets\arg\min_{g\in K}D_g$ \EndIf
  \If{$U_K\ne\varnothing$}
    \If{$s_K\notin U_K$} \State $s_K\gets\arg\max_{g\in U_K}\sigma_g$ \EndIf
    \State $h_K\gets s_K$
  \Else
    \State $s_K\gets\bot$; $h_K\gets c_K$
  \EndIf
\EndFor
\AlgNote{Allocate and redistribute link budgets}
\ForAll{active directed links $e$}
  \State $K\gets$ component containing $\mathcal{G}_e$
  \State $\mathcal{R}_e\gets\{g:\exists q\in\mathcal{Q}_{e,g},\ q\sim\mathcal{B}_g\}$
  \State $a_{e,g}\gets1$ for all $g\in\mathcal{G}_e$
  \If{$(U_K=\varnothing\text{ or }\mathcal{R}_e\ne\varnothing)$ and $h_K\in\mathcal{G}_e$}
    \State $a_{e,h_K}\gets\alpha$
  \EndIf
  \State $b_{e,g}\gets\widetilde b_{e,g}\gets\widehat C_e a_{e,g}/\sum_h a_{e,h}$
  \If{$\mathcal{R}_e=\varnothing$} \State \textbf{continue} \EndIf
  \State $D_e\gets0$
  \ForAll{$g\in\mathcal{G}_e\setminus\mathcal{R}_e$}
    \State $H_g\gets R_g-t-\delta$
    \If{$V_g$ and $\mathcal{B}_g=\varnothing$ and $D_g<H_g<+\infty$}
      \State $d\gets[\widetilde b_{e,g}-\max(L_{e,g}/H_g,\rho\widetilde b_{e,g})]^+$
      \State $b_{e,g}\gets b_{e,g}-d$; $D_e\gets D_e+d$
    \EndIf
  \EndFor
  \ForAll{$g\in\mathcal{R}_e$} \State $b_{e,g}\gets b_{e,g}+D_e/|\mathcal{R}_e|$ \EndFor
\EndFor
\State \Return $\mathbf{b}$
\end{algorithmic}
\end{algorithm}

\textbf{Identifying blocking dependencies.}
The runtime reports collective membership, successor dependencies, committed GPU work, and communication progress. Delivered records give each consumer's readiness:
\begin{equation}
R_{g,i}=\max\{T_i^{\mathrm{queue}},T_{g,i}^{\mathrm{own}},T_{g,i}^{\mathrm{release}}\},
\label{eq:ready}
\end{equation}
from the GPU queue tail, predecessor completion, and operation release. Let $R_g$ be the earliest readiness among unfinished consumers. Group validity $V_g$ checks membership, a common successor, fresh GPU records, and outstanding-data evidence. For consumer $G_i$, let $\mathcal I_{g,i}$ and $\mathcal O_{g,i}$ be the required and observed incident-flow identities. The blocked set $\mathcal B_g$ contains consumers with $R_{g,i}\le t$, $\mathcal O_{g,i}=\mathcal I_{g,i}$, fresh pending-transfer records, and unfinished communication. Groups without valid consumer metadata have $\mathcal B_g=\varnothing$ and cannot donate computation slack.

Let $u_q,o_q$ be unsent and unacknowledged bytes. Accounting for packet payload $m$ and overhead $h_{\mathrm{pkt}}$ with $h(v)=v+h_{\mathrm{pkt}}\lceil v/m\rceil$, remaining wire work is
\begin{equation}
L_{e,g}=8\sum_{q\in\mathcal Q_{e,g}}h(u_q)
+8\sum_{\substack{q\in\mathcal F_g:\,e\in\mathcal P_q}}h(o_q),
\label{eq:wirework}
\end{equation}
where $\mathcal F_g$ includes observed flows awaiting their final ACK. Outstanding bytes remain charged to their paths until acknowledgment.

\textbf{Coordinating collective priorities.}
Equal-share budgets provide a reference drain estimate:
\begin{equation}
b^0_{e,g}=\frac{\widehat C_e}{|\mathcal G_e|},\qquad
D_g=\max_{e:g\in\mathcal G_e}\frac{L_{e,g}}{b^0_{e,g}}.
\label{eq:drain}
\end{equation}
For groups with blocked consumers, the priority score
\begin{equation}
\sigma_g=\frac{|\mathcal B_g|}{\max(\tau_0,D_g)}
\label{eq:score}
\end{equation}
measures waiting consumers per unit of estimated drain time; $\tau_0$ bounds the score near completion. The maximum in $D_g$ captures the most restrictive path segment, so the score reflects both the remaining collective work and the consumers it can release.

Connect collectives that share a directed link. Within each connected component $K$, a blocked-consumer winner $s_K$ retains priority while eligible; otherwise the largest $\sigma_g$ wins. Independently, a continuity group $c_K$ persists while active, or is replaced by the group with smallest $D_g$. It supplies priority when no blocked group exists. Ties use collective IDs. One collective can contend with different groups along its path. Independent link priorities could split service across competing dependencies. A component-wide winner gives those links a common direction, while each budget remains specific to the groups traversing that link.

Write $q\sim\mathcal B_g$ when either endpoint of $q$ belongs to $\mathcal B_g$, and let $\mathcal R_e$ contain groups with such a QP on $e$. With blocked consumers present, the winner receives weight $\alpha\ge1$ on its links where $\mathcal R_e\ne\varnothing$; other links retain equal shares. With no blocked group, the continuity winner receives that weight along its paths. Initial budgets are
\begin{equation}
\widetilde b_{e,g}=\widehat C_e\frac{a_{e,g}}{\sum_{h\in\mathcal G_e}a_{e,h}},
\qquad a_{e,g}\in\{1,\alpha\}.
\label{eq:groupbudget}
\end{equation}

\begin{algorithm}[t]
\caption{Weighted progressive filling}
\label{alg:fill}
\small
\begin{algorithmic}[1]
\Require Constraint rows $A$, capacities $\mathbf c$, weights $\mathbf w>0$
\State $\mathbf x\gets\mathbf0$; $U\gets\mathcal Q$
\While{$U\ne\varnothing$}
  \State $s_j\gets\sum_{q\in U}A_{jq}w_q$ for each row $j$
  \State $\Delta\gets\min_{j:s_j>0}\{[c_j-(A\mathbf x)_j]/s_j\}$
  \ForAll{$q\in U$} \State $x_q\gets x_q+w_q\Delta$ \EndFor
  \State $J\gets\{j:(A\mathbf x)_j=c_j\}$
  \State $U\gets U\setminus\{q:\exists j\in J,\ A_{jq}=1\}$
\EndWhile
\State \Return $\mathbf x$
\end{algorithmic}
\end{algorithm}

\textbf{Reassigning computation slack.}
A donor has valid state, no blocked consumer, finite $R_g$, and enough time before its earliest consumer is ready:
\begin{equation}
H_g=R_g-t-\delta>D_g,
\label{eq:donor}
\end{equation}
where $\delta$ reserves a guard interval. Using the earliest unfinished consumer ends the window when any consumer in the group can first need the communication. Its retained budget and donation are
\begin{align}
k_{e,g}&=\max\{L_{e,g}/H_g,\rho\widetilde b_{e,g}\},\label{eq:keep}\\
d_{e,g}&=[\widetilde b_{e,g}-k_{e,g}]^+,\qquad 0<\rho\le1.
\label{eq:donation}
\end{align}
The terms in $k_{e,g}$ cover the observed work over the available horizon and preserve a service floor. Algorithm~\ref{alg:budgets} pools donations on each link and shares them equally among $\mathcal R_e$. The horizon condition governs this reassignment after initial priority allocation. A recipient need not be the component winner: eligibility follows the blocked consumers served on that link. Priority coordinates collective progress across a component, while slack reassignment uses computation overlap to assist blocking transfers.

\begin{algorithm}[t]
\caption{Joint ACK/INT rate update}
\label{alg:ack}
\small
\begin{algorithmic}[1]
\Require ACK/INT $a$, cache $\Lambda$, last refresh $t_0$
\AlgEvent{On arrival of an ACK for QP $q$:}
\State $t\gets\operatorname{Now}()$; $\operatorname{RecordProgress}(q,a)$
\State $\mathcal N\gets\operatorname{UpdateNetwork}(q,a.I,t)$
\If{cache uninitialized or $t-t_0\ge\Delta_{\mathrm{ref}}$}
  \State $S\gets\operatorname{ReadDelivered}(t)$
  \State $\widehat{\mathbf C}\gets\operatorname{LinkBudgets}(\mathcal N,t)$ \Comment{Eq.~\ref{eq:effectivecapacity}}
  \State $(V,R,\mathcal B)\gets\operatorname{Classify}(S,t)$
  \State $\mathbf b\gets\operatorname{GroupBudgets}(S,\widehat{\mathbf C},V,R,\mathcal B,t)$
  \State $(A,\mathbf c,\mathbf w)\gets\operatorname{BuildConstraints}(S,\mathbf b,\mathcal B)$
  \State $\mathbf x\gets\operatorname{ProgressiveFill}(A,\mathbf c,\mathbf w)$ \Comment{Alg.~\ref{alg:fill}}
  \State $\Lambda\gets\varnothing$
  \ForAll{$j\in\mathcal Q$}
    \State $\Lambda[j]\gets(\max(r_{\min},\lfloor x_j\rfloor),t+T_{\mathrm{life}})$
  \EndFor
  \State $t_0\gets t$; $J\gets\operatorname{LiveManagedQPs}()$
\Else
  \State $J\gets\{q\}$
\EndIf
\ForAll{unfinished $j\in J$}
  \If{$j\notin\Lambda$ or $t>\Lambda[j].\mathrm{expiry}$}
    \State \textbf{continue} \Comment{Keep the applied rate}
  \EndIf
  \State $\operatorname{ApplyPacing}(j,\max\{r_{\min},\min(\Lambda[j].\mathrm{rate},P_j)\})$
\EndFor
\end{algorithmic}
\end{algorithm}

\textbf{Proposition 1 (budget conservation).}
In continuous arithmetic, Algorithm~\ref{alg:budgets} satisfies $\sum_g b_{e,g}=\widehat C_e\le C_e$ and $b_{e,g}\ge\rho\widetilde b_{e,g}>0$.
\emph{Proof.} Initial shares sum to $\widehat C_e$. A donor releases at most $(1-\rho)\widetilde b_{e,g}$, and recipients receive exactly the released sum. Downward rounding preserves the capacity inequality. \hfill$\square$

\subsection{QP Allocation and Rate Execution}
\label{sec:qpallocation}
\label{sec:rateupdate}

\textbf{Allocating within collective budgets.}
QP weights combine remaining work with blocked-consumer priority:
\begin{equation}
w_q=\begin{cases}
\max(1,u_q),&q\sim\mathcal B_{g(q)},\\
\max(1,u_q)/\beta,&\text{otherwise},
\end{cases}\qquad\beta\ge1.
\label{eq:qpweight}
\end{equation}
For equally classified QPs, proportional rates equalize remaining serialization times until a path constraint binds; $\beta$ favors QPs associated with blocked consumers.

Write the link--group and endpoint constraints as $A\mathbf x\le\mathbf c$. A link--group row selects $\mathcal Q_{e,g}$ and has capacity $b_{e,g}$; an endpoint row selects one QP and has capacity $P_q$. IACC applies weighted progressive filling to obtain
\begin{equation}
\underset{\mathbf x\in\mathcal X(\mathbf b)}{\operatorname{lexmax}}
\ \operatorname{sort}_{\uparrow}\{x_q/w_q:q\in\mathcal Q\}.
\label{eq:lexmax}
\end{equation}
For active QPs $U$, define $s_j=\sum_{q\in U}A_{jq}w_q$. Each fill step increases their rates at slope $w_q$ by the largest feasible common increment:
\begin{equation}
\Delta=\min_{j:s_j>0}\frac{c_j-(A\mathbf x)_j}{s_j}.
\label{eq:fillstep}
\end{equation}
Algorithm~\ref{alg:fill} freezes QPs touching a saturated constraint and continues with the rest.

\textbf{Proposition 2 (fixed-snapshot allocation).}
For fixed budgets and positive weights, Algorithm~\ref{alg:fill} returns a feasible weighted max-min allocation in at most $n=|\mathcal Q|$ iterations.
\emph{Proof.} Equation~\eqref{eq:fillstep} preserves feasibility and freezes at least one QP per step. Each frozen QP reaches an endpoint cap or shares a saturated budget with QPs of no larger normalized rate, establishing weighted max-min fairness. \hfill$\square$
With $M=\sum_q|\mathcal P_q|$, sparse scans give $O(n(n+M))$ time and $O(n+M)$ storage.

\textbf{Applying the rate decision.}
Algorithm~\ref{alg:ack} updates network and progress records on each ACK. At a refresh it computes link budgets, collective budgets, and QP rates, replaces the cache $\Lambda$, and updates matching unfinished QPs. Between refreshes, only the current ACK's QP is considered. With quantized target $\bar x_q=\max\{r_{\min},\lfloor x_q\rfloor\}$, execution follows
\begin{equation}
r_q(t^+)=
\begin{cases}
\max\{r_{\min},\min(\bar x_q,P_q)\},&\Lambda[q]\text{ valid},\\
r_q(t^-),&\text{otherwise}.
\end{cases}
\label{eq:directrate}
\end{equation}
A new QP starts at $P_q$; missing or expired grants retain the applied rate. Network and consumer state thus determine one directly applied absolute rate. Quantization adds at most $n_er_{\min}$ above the continuous link budget, where $n_e$ counts managed QPs on $e$. The allocation properties apply to each fixed snapshot; observation delay, caching, and in-flight packets also affect execution.

%% file: sections/05_evaluation.tex
\section{Evaluation}
\label{sec:evaluation}
\raggedbottom

\begin{figure}[!t]
\centering
\includegraphics[width=\columnwidth]{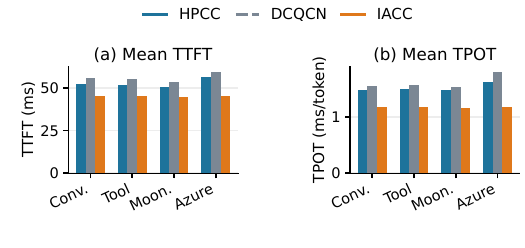}
\caption{Mean TTFT and per-request TPOT at $1.0\times$ arrival density. Four workloads are shown; BurstGPT is included in the density sweep and aggregate statistics. Lower is better.}
\label{fig:evalworkloads}
\end{figure}
\begin{figure*}[!t]
\centering
\includegraphics[width=\textwidth]{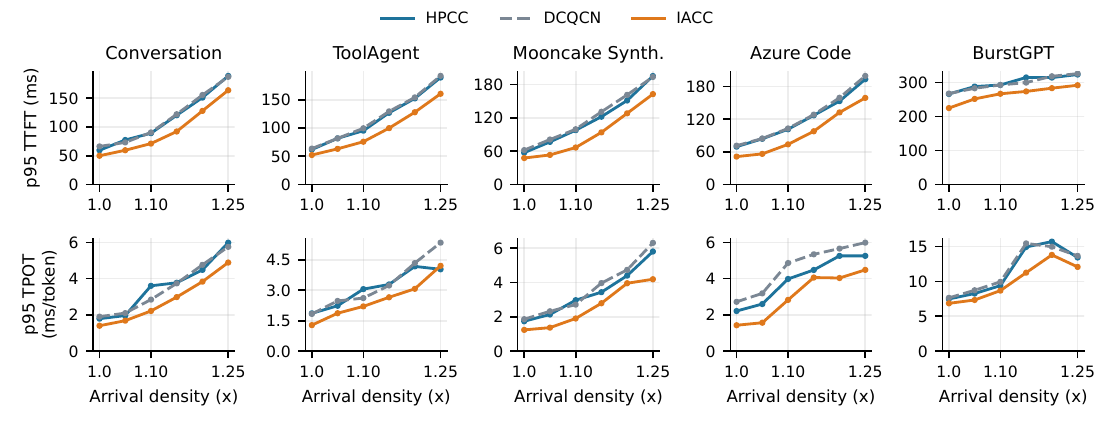}
\caption{p95 TTFT (top) and per-request p95 TPOT (bottom) versus arrival density. Each column corresponds to one workload; vertical scales differ across panels. Methods share the same inputs and measurement window at each setting.}
\label{fig:evaldensity}
\end{figure*}

We evaluate how IACC affects token latency and GPU waiting. We first compare end-to-end performance across five workloads, then relate the gains to communication and computation progress. Finally, we isolate the effects of consumer-state information and intra-collective QP weighting.


\subsection{Experimental Setup}
\label{sec:evalsetup}

\textbf{Simulation environment.}
We evaluate IACC using SimAI's packet-level ns-3 backend~\cite{simai}. The cluster contains 16 GPUs in a disaggregated prefill/decode deployment: 12 prefill GPUs form three EP groups, and four decode GPUs form one EP group, all with EP=4. Four leaf switches connect to one spine, with four GPUs attached to each leaf. Each EP group spans all four leaves, creating contention among concurrent collectives and KV transfers. GPU--leaf and leaf--spine links provide 50 and 25\,Gb/s, respectively, with 2\,$\mu$s propagation delay per link.\footnote{Internally, the simulator scales time by $1000\times$ and reduces link rates by the same factor. Reported latencies use the reference scale; time-series axes retain simulation time.}

\textbf{Workloads.}
We use five trace-guided inference workloads: Conversation, ToolAgent, and Mooncake Synthetic from Mooncake~\cite{mooncake_traces}, Azure Code~\cite{azure_traces}, and BurstGPT~\cite{burstgpt_traces}. Trace windows provide arrival patterns and input/output-length ordering, which are mapped to synthetic prefill/decode templates with explicit computation and communication dependencies. Each request has 8, 12, or 16 decode iterations. The uncompressed arrival sequence of each selected trace window defines $1.0\times$ density. Arrival density is varied from $1.0\times$ to $1.25\times$ in $0.05\times$ steps, with additional tests at $1.5\times$, $1.75\times$, and $2.0\times$. Increasing density compresses request interarrival times while preserving request order and per-request work. All traffic comes from the inference workloads.

\textbf{Comparisons and measurement.}
We compare IACC with HPCC~\cite{hpcc} and DCQCN~\cite{dcqcn}. Methods use identical request inputs, placement, and execution dependencies on the same topology. Each workload--density setting uses one trace window and a paired run per method. Runs include 60 simulation seconds of warm-up and a 600-second measurement window, followed by draining. Statistics cover all requests arriving during the measurement window, including those completing after its end.

\textbf{Metrics.}
Our primary metrics are mean and p95 time to first token (TTFT) and time per output token (TPOT). For request arrival time $a_r$ and token-ready times $t_{r,1},\ldots,t_{r,N_r}$, TTFT is $t_{r,1}-a_r$ and TPOT is $(t_{r,N_r}-t_{r,1})/(N_r-1)$. TPOT is computed per request before aggregation. A token becomes ready when its decode iteration's required communication and computation complete. We also report request completion time (RCT), $t_{r,N_r}-a_r$, and the fraction of requests satisfying both TTFT$\le55$\,ms and TPOT$\le1.7$\,ms/token.

To explain the gains, we measure collective completion time, KV handoff latency, GPU waiting, and utilization. Collective completion spans first issue to last-rank completion; KV handoff spans producer readiness to arrival of all required data. GPU waiting measures idle time while subsequent computation remains. For each GPU, task intervals begin at request arrival for prefill or complete KV arrival for decode, and end at the last positive-duration computation. Overlapping intervals are merged. Utilization is the fraction of time spent computing within these intervals, averaged across GPUs.

\begin{figure*}[!t]
\centering
\includegraphics[width=\textwidth]{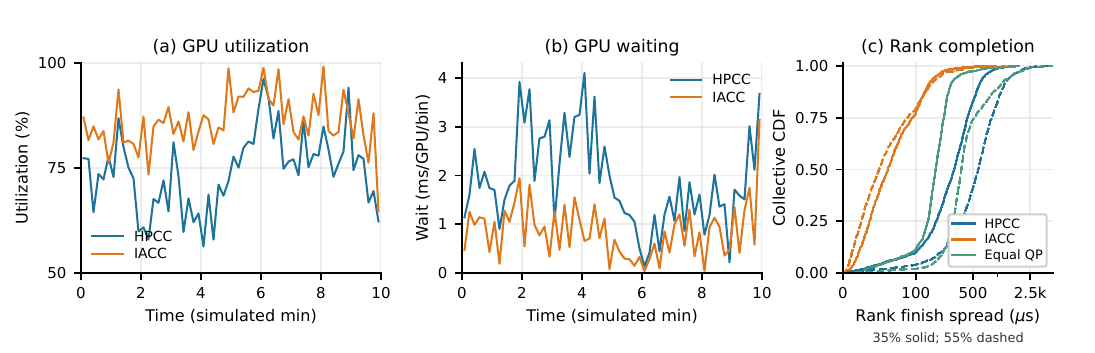}
\caption{Communication and GPU progress in Conversation at $1.0\times$. (a)--(b) Original workload: GPU utilization and GPU waiting in 10-simulation-second bins. Utilization uses busy/(busy+waiting) within task intervals. (c) Controlled rank imbalance: solid/dashed CDFs denote 35\%/55\% hot-pair shares. The spread axis is linear through 100\,$\mu$s and logarithmic thereafter.}
\label{fig:evalprogress}
\end{figure*}

\subsection{End-to-End Performance}
\label{sec:evale2e}

\textbf{Token latency across workloads.}
Figure~\ref{fig:evalworkloads} compares Conversation, ToolAgent, Mooncake Synthetic, and Azure Code at $1.0\times$ arrival density. Across all five workloads, including BurstGPT, IACC reduces mean TTFT relative to HPCC by 6.32\%--19.41\%, mean TPOT by 7.02\%--27.33\%, and mean RCT by 6.45\%--21.47\%. Azure Code shows the largest reductions in mean TTFT and TPOT, while the bursty BurstGPT window exhibits smaller improvements.

\textbf{Increasing arrival pressure.}
Figure~\ref{fig:evaldensity} shows p95 TTFT and TPOT as arrival density increases. Relative to HPCC, IACC reduces mean and p95 TTFT in all 30 workload--density settings, by 5.81\%--28.02\% and 8.85\%--33.00\%, respectively. Mean TPOT improves in 27 settings and p95 TPOT in 29; mean RCT falls by 6.45\%--30.11\%. For ToolAgent at $1.25\times$, p95 TPOT is 4.25\% higher than HPCC. Overall, IACC improves token latency in most settings and reduces request completion time as arrival pressure increases.

\textbf{Joint service quality.}
We apply the same TTFT and TPOT budgets to every setting. At $1.0\times$, all IACC requests meet both budgets for Conversation, ToolAgent, Mooncake Synthetic, and Azure Code, compared with 42.86\%--75.00\% under HPCC. At $1.10\times$, IACC achieves 38.71\%--67.74\% for these workloads, versus 0\%--9.68\% for HPCC. BurstGPT has a zero joint pass rate at baseline density under all three methods. At densities of $1.5\times$ and above, all five workloads have zero joint pass rate. At $2.0\times$, 75.44\%--84.62\% of HPCC's measurement-cohort requests and 71.93\%--84.62\% of IACC's remain unfinished at the window boundary. These requests complete during draining and remain included in the latency statistics. The results characterize performance under heavy load over the measurement window rather than a maximum sustainable arrival rate.

\subsection{Understanding the Gains}
\label{sec:evalgains}

\textbf{From communication progress to token latency.}
We use Conversation at $1.0\times$ to examine the stages behind the aggregate gains. In this illustrative case, all 28 matched request IDs have lower TTFT, TPOT, and RCT under IACC. Figure~\ref{fig:evalprogress}(a)--(b) tracks GPU utilization and GPU waiting over the common measurement window.

Table~\ref{tab:evalttft} decomposes mean TTFT into arrival-to-KV-readiness at the producer, KV handoff, and KV-arrival-to-first-token. IACC reduces all three components. Prefill and decode collective completion times fall by 19.88\% and 21.55\%, while KV handoff falls by 13.90\%. The arrival-to-prefill-readiness interval combines queueing, computation, and prefill communication, linking the communication changes to progress through the complete execution path.

\begin{table}[t]
\centering
\caption{Mean TTFT breakdown for Conversation at $1.0\times$ (ms). Prefill includes arrival-to-producer-KV-readiness; Post-KV ends at the first token.}
\label{tab:evalttft}
\small
\setlength{\tabcolsep}{4pt}
\begin{tabular}{@{}lrrrr@{}}
\toprule
Method & Prefill & KV & Post-KV & TTFT \\
\midrule
HPCC & 43.849 & 7.224 & 1.161 & 52.235 \\
IACC & 38.072 & 6.220 & 0.648 & 44.940 \\
\bottomrule
\end{tabular}
\end{table}

\textbf{Reducing idle intervals.}
GPU waiting drops by 54.86\%, and utilization rises from 73.08\% to 85.65\%. The lower token latencies therefore accompany less idle time during active execution, with nearly unchanged computation in the observation window. This connects the request-level gains to shorter waits along the communication--computation dependency chain.

\begin{figure*}[t]
\centering
\includegraphics[width=\textwidth]{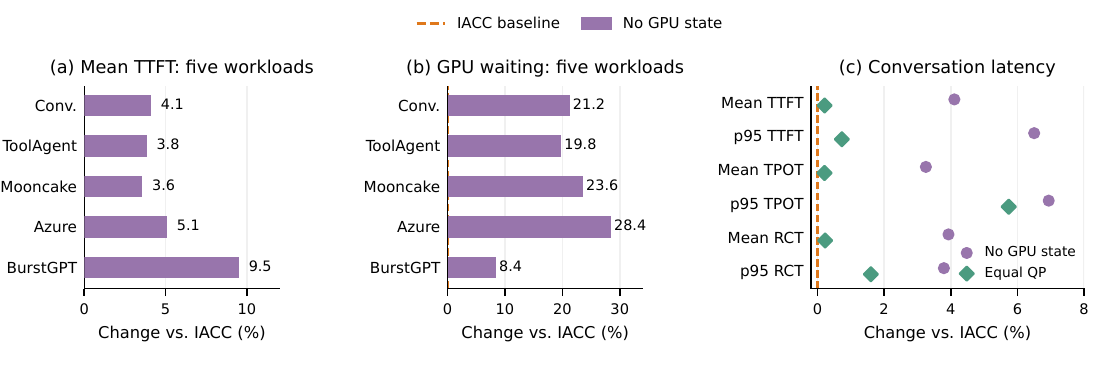}
\caption{Ablation at $1.0\times$ arrival density. (a)--(b) Changes in mean TTFT and GPU waiting when consumer state is disabled across five workloads. (c) Mean and p95 request latencies for Conversation under NG and EQ. Values are percentage changes relative to full IACC; the dashed zero line denotes full IACC. Positive values indicate higher latency or waiting.}
\label{fig:evalablation}
\end{figure*}

\textbf{Controlled rank imbalance.}
Token routing produces uneven expert demand and All-to-All traffic across ranks. Lina reports that the most popular expert receives 4.02 and 5.56 times as many tokens as the least popular expert in its four- and sixteen-expert inference tasks~\cite{lina}. Tutel also identifies dynamically changing expert loads~\cite{tutel}. We use controlled traffic skew to examine how uneven rank progress affects consumer waiting and token latency.

For Conversation at $1.0\times$, we preserve request arrivals, computation, KV traffic, and each collective's total payload, and vary only the distribution across rank pairs. EP=4 produces 12 directed flows. The two reciprocal flows of a hot rank pair carry 35\% or 55\% of the payload; the other ten flows share the remainder equally. The hot pair rotates across communication rounds. A uniform distribution assigns 16.7\% to a pair, so the two settings concentrate 2.1 and 3.3 times the uniform share, respectively. Table~\ref{tab:evalrank} also includes NG and EQ, defined in Section~\ref{sec:evalablation}.

Figure~\ref{fig:evalprogress}(c) shows rank-finish spread, the difference between the earliest and latest rank completion times. We also measure local consumer waiting by merging GPU idle intervals after the earliest rank finishes while the consumer still awaits communication. At 35\% skew, IACC reduces mean spread from 325.8 to 75.3\,$\mu$s; at 55\%, it falls from 659.1 to 72.8\,$\mu$s. Local consumer waiting falls from 66.86 to 15.04\,ms/GPU and from 113.54 to 12.21\,ms/GPU, respectively.

\begin{table}[!t]
\centering
\caption{Latency and rank-finish spread under controlled traffic skew. TTFT is in ms, TPOT in ms/token, and spread in $\mu$s. Columns report means unless marked p95.}
\label{tab:evalrank}
\small
\setlength{\tabcolsep}{3pt}
\begin{tabular}{@{}llrrrr@{}}
\toprule
Share & Method & TTFT & TPOT & p95 TPOT & Spread \\
\midrule
35\% & HPCC & 96.00 & 2.448 & 3.515 & 325.8 \\
 & IACC & 81.65 & 2.078 & 2.737 & 75.3 \\
 & NG & 112.59 & 1.743 & 2.348 & 46.5 \\
 & EQ & 76.03 & 2.043 & 3.323 & 193.0 \\
\midrule
55\% & HPCC & 134.32 & 3.515 & 4.903 & 659.1 \\
 & IACC & 127.00 & 2.986 & 4.502 & 72.8 \\
 & NG & 224.71 & 3.627 & 5.206 & 51.6 \\
 & EQ & 129.48 & 3.172 & 4.667 & 511.2 \\
\bottomrule
\end{tabular}
\end{table}

Table~\ref{tab:evalrank} relates rank progress to end-to-end latency. At 35\% skew, IACC reduces mean TTFT, TPOT, and RCT relative to HPCC by 14.95\%, 15.10\%, and 14.69\%; their p95 values fall by 16.69\%, 22.13\%, and 18.22\%. At 55\% skew, mean TTFT and TPOT fall by 5.45\% and 15.06\%. These results connect more concentrated rank completion and shorter local waits to improvements in token latency.

Relative to EQ, full IACC reduces p95 TPOT by 17.63\%, mean rank-finish spread by 60.97\%, and local consumer waiting by 58.45\% at 35\% skew. Mean TTFT, mean TPOT, and total GPU waiting increase by 7.40\%, 1.71\%, and 9.82\%, respectively. At 55\% skew, mean TTFT and TPOT fall by 1.91\% and 5.86\%. The effect of intra-group weighting therefore depends on the traffic distribution and the execution stages exposed to communication delay.

\subsection{Ablation}
\label{sec:evalablation}

We separately examine consumer-state awareness and intra-group weighting. All methods use the same request inputs, computation dependencies, and network configuration, with only the corresponding component disabled.

\textbf{Consumer-state awareness.}
\emph{No GPU state} (NG) removes GPU busy, queue, and consumer-urgency information while retaining collective progress and link constraints. Figure~\ref{fig:evalablation}(a)--(b) compares the five workloads at $1.0\times$ arrival density. Disabling consumer state increases mean TTFT by 3.57\%--9.51\% and GPU waiting by 8.39\%--28.42\%. Consumer information helps the rate decision identify communication that limits computation, reducing waiting and request latency.

\textbf{Intra-group weighting.}
\emph{Equal QP} (EQ) retains consumer state, group budgets, and link constraints, but sets all intra-group QP weights to one. This isolates the effect of nonuniform weighting. Across the five original workloads, EQ changes mean TTFT by $-0.04\%$ to $+0.24\%$ and mean TPOT by $-6.56\%$ to $+0.89\%$ relative to full IACC. The effect on mean latency is modest and workload dependent.

Figure~\ref{fig:evalablation}(c) compares the six request-latency metrics for Conversation at $1.0\times$. Full IACC has lower values for all six latency metrics than both variants. Relative to NG, mean TTFT, TPOT, and RCT fall by 3.95\%, 3.15\%, and 3.78\%. Relative to EQ, the reductions are 0.22\%, 0.21\%, and 0.23\% for the means, and 0.73\%, 5.43\%, and 1.58\% for their p95 values. The main benefit of weighting in this workload appears in the TPOT tail.

In this case, GPU waiting is 50.88\,ms/GPU under full IACC, slightly above EQ's 49.92\,ms/GPU; utilization is 85.65\% and 85.88\%, respectively. Together with the imbalance experiments in Section~\ref{sec:evalgains}, these results show a clearer contribution from consumer state to mean request latency. Intra-group weighting changes communication completion order and its effect on subsequent computation.

%% file: sections/06_related_work.tex
\section{Related Work}
\label{sec:related}

\textbf{Datacenter congestion control.}
Datacenter congestion-control designs use different signals and objectives. DCTCP and DCQCN use ECN feedback, while TIMELY and Swift infer congestion from delay~\cite{dctcp,dcqcn,timely,swift}. HPCC and PowerTCP use in-network telemetry to estimate link utilization and react to congestion~\cite{hpcc,powertcp}; PCC selects rate actions based on observed end-to-end performance~\cite{pcc}. Other designs target latency or flow completion: HULL reserves bandwidth headroom to limit queueing~\cite{hull}, pFabric prioritizes flows by their completion requirements~\cite{pfabric}, and D$^3$ allocates rates using flow deadlines~\cite{d3}. Together, these designs establish feedback-driven rate control and flow-level prioritization. IACC extends this line of work by incorporating collective dependencies and GPU readiness when assigning service among competing QPs.

\textbf{Application-aware bandwidth allocation.}
D$^2$TCP incorporates flow deadlines into congestion control~\cite{d2tcp}. Coflow research represents collections of related flows by their shared application-level completion objective~\cite{coflow}; Varys uses this abstraction to schedule communication stages, and Aalo develops a non-clairvoyant coflow scheduler~\cite{varys,aalo}. MLTCP uses iteration progress to coordinate communication across distributed training jobs~\cite{mltcp}, while ByteScheduler schedules and partitions DNN-training transfers while preserving framework dependencies~\cite{bytescheduler}. These systems demonstrate that application structure can guide network service. IACC focuses on inference-time consumer readiness: it identifies communication that blocks GPU execution and computation intervals that can absorb delayed delivery, then uses this state to reallocate bandwidth among collectives inside the ACK/INT feedback path.

\textbf{MoE communication optimization.}
MoE systems optimize expert placement, load balance, and communication overlap. DeepSpeed-MoE and Tutel provide scalable MoE execution, while FasterMoE and Lina target communication and expert-load inefficiencies in distributed MoE workloads~\cite{deepspeed_moe,tutel,fastermoe,lina}. MegaBlocks improves sparse expert computation, and MegaScale-Infer uses disaggregation and pipelining for MoE serving~\cite{megablocks,megascale}. These systems organize computation and communication at the model or serving-system level. IACC addresses a complementary transport question: given concurrent collectives and their runtime dependencies, how should shared-link service be divided to advance the GPU computations waiting for communication results?

\textbf{LLM serving systems.}
Serving systems also schedule work to balance throughput and latency. Orca introduces iteration-level scheduling for autoregressive generation, and Sarathi-Serve uses chunked prefills to reduce interference between prefill and decode iterations~\cite{orca,sarathi}. DistServe and Splitwise separate prefill and decode onto different GPUs or machines to manage their distinct resource demands~\cite{distserve,splitwise}. These systems primarily shape request batching, phase placement, or GPU allocation. IACC instead controls the rates of concurrent communication flows based on network feedback and the readiness of their GPU consumers, and can complement these serving-level policies.

%% file: sections/07_conclusion.tex
\section{Conclusion}
\label{sec:conclusion}

Distributed MoE inference couples network service to collective dependencies and GPU readiness, so link utilization alone cannot characterize communication's effect on request progress. IACC incorporates these execution signals into ACK/INT-based QP-rate allocation; its fixed-snapshot analysis establishes budget conservation and weighted max-min allocation. Across five trace-guided workloads, IACC reduces mean TTFT, TPOT, and RCT by 6.32\%--19.41\%, 7.02\%--27.33\%, and 6.45\%--21.47\%, respectively, relative to HPCC at baseline arrival density. At arrival densities through $1.25\times$, mean and p95 TTFT are lower in all 30 tested settings. For Conversation, GPU waiting falls by 54.86\%, while utilization rises from 73.08\% to 85.65\%. These results show how aligning network service with consumer readiness can improve end-to-end progress in the evaluated workloads.